%% file: HIN-14-001_temp.tex
\pdfoutput=1

\documentclass[11pt,twoside,a4paper,cmspaper,final,collab]{cms-tdr}
\def\svnVersion{354120}\def\cmsCernNoTag{CERN-PH-EP/2015-334}\def\cmsCernDate{\today}\def\cmsMessage{Published in the European Physical Journal C as \href{http://dx.doi.org/10.1140/epjc/s10052-016-4205-7}{\doi{10.1140/epjc/s10052-016-4205-7.}}}

\begin{document}\cmsNoteHeader{HIN-14-001}

\hyphenation{had-ron-i-za-tion}
\hyphenation{cal-or-i-me-ter}
\hyphenation{de-vices}
\RCS$Revision: 344861 $
\RCS$HeadURL: svn+ssh://svn.cern.ch/reps/tdr2/papers/HIN-14-001/trunk/HIN-14-001.tex $
\RCS$Id: HIN-14-001.tex 344861 2016-05-30 19:10:18Z velkovsk $

\providecommand {\pp} {\ensuremath{\Pp\Pp}\xspace}
\providecommand {\pbarp} {\ensuremath{\Pp\PAp}\xspace}
\providecommand {\Pb}  {\text{Pb}\xspace}
\providecommand {\PbPb}  {\text{PbPb}\xspace}
\providecommand {\pPb}  {\ensuremath{{\Pp\mathrm{Pb}}}\xspace}
\providecommand {\sqrtsNN}  {\ensuremath{\sqrt{s_{_\mathrm{NN}}}}\xspace}
\providecommand{\HIJING}{\textsc{hijing}\xspace}
\providecommand{\PYTHIAHIJING}{\textsc{pythia+hijing}\xspace}
\providecommand{\POWHEGPYTHIA}{\textsc{powheg+pythia}\xspace}
\newlength\cmsFigWidth
\ifthenelse{\boolean{cms@external}}{\setlength\cmsFigWidth{0.99\columnwidth}}{\setlength\cmsFigWidth{0.6\textwidth}}
\ifthenelse{\boolean{cms@external}}{\providecommand{\cmsLeft}{Top}\xspace}{\providecommand{\cmsLeft}{Left}\xspace}
\ifthenelse{\boolean{cms@external}}{\providecommand{\cmsRight}{Bottom}\xspace}{\providecommand{\cmsRight}{Right}\xspace}
\ifthenelse{\boolean{cms@external}}{\providecommand{\cmsleft}{top }\xspace}{\providecommand{\cmsleft}{left }\xspace}
\ifthenelse{\boolean{cms@external}}{\providecommand{\cmsright}{bottom }\xspace}{\providecommand{\cmsright}{right }\xspace}
\cmsNoteHeader{HIN-14-001}
\title{Measurement of inclusive jet production and nuclear modifications in pPb collisions at $\sqrtsNN=5.02\TeV$}

\date{\today}

\abstract{
Inclusive jet production in pPb collisions at a nucleon-nucleon (NN) center-of-mass energy of $\sqrtsNN=5.02\TeV$ is studied with the CMS detector at the LHC. A data sample corresponding to an integrated luminosity of 30.1\nbinv is analyzed.
The jet transverse momentum spectra are studied in seven  pseudorapidity intervals covering  the range $ -2.0 <\eta_\mathrm{CM}< 1.5$ in  the NN center-of-mass frame. The jet production yields at forward and backward  pseudorapidity are compared and  no significant asymmetry about $\eta_\mathrm{CM} = 0$  is observed in the measured kinematic range. The measurements in the pPb system are compared to reference jet spectra obtained by extrapolation from previous measurements in pp collisions  at  $\sqrt{s}=7\TeV$. In all pseudorapidity ranges, nuclear modifications in inclusive jet production are found to be small, as predicted by next-to-leading order perturbative QCD calculations that incorporate nuclear effects in the parton distribution functions.
}

\hypersetup{%
pdfauthor={CMS Collaboration},%
pdftitle={Measurement of inclusive jet production and nuclear modifications in pPb collisions at sqrt(s[NN]) = 5.02  TeV},%
pdfsubject={CMS},%
pdfkeywords={CMS, heavy ion physics, jet nuclear modification}}

\maketitle

\section{Introduction}
\label{sec:introduction}

Jet measurements play an important role in the study of the quark gluon plasma (QGP) produced in relativistic heavy ion collisions. A key observable in these studies is the phenomenon of jet quenching~\cite{Bjorken:1982tu,2010LanB...23..471D,CasalderreySolana:2007zz,Appel:1985dq,Blaizot:1986ma,Qiu:2003pm}, in which the partons produced in hard scattering lose energy through gluon radiation and elastic scattering in the hot and dense partonic medium. Jet quenching was first observed at RHIC through measurements of high transverse momentum (\pt) hadrons~\cite{Adcox:2001jp} and dihadron
 correlations~\cite{STAR:2003}. At the LHC, this phenomenon was observed more directly as dijet momentum imbalance~\cite{Aad:2010bu,Chatrchyan:2011sx} and photon-jet energy imbalance~\cite{Chatrchyan:2012gt} in \PbPb collisions. An important ingredient in understanding how the presence of a hot  QCD medium affects the jets is the comparison to reference measurements from collision systems that are not expected to produce the QGP.
 Most often, \pp collisions at the same center-of-mass energy are used as a reference. Modifications in  jet yields~\cite{Aad:2014bxa,Adam:2015ewa}, shapes~\cite{Chatrchyan:2013kwa}, and fragmentation patterns~\cite{Aad:2014wha,Chatrchyan:2014ava} in \PbPb collisions have been  found  in comparison to expectations based on  \pp measurements. These modifications are found to depend on the overlap between the colliding nuclei, and are  largest in the most central (i.e., largest overlap) \PbPb collisions.

The interpretation of the jet modification results in nucleus-nucleus collisions and the understanding of their relation to the properties of the QGP requires detailed knowledge of all nuclear effects that could influence the comparisons with the \pp system. Nuclear modifications may already be present at the initial state of the collisions, independently of QGP formation. Such modifications are collectively referred to as cold nuclear matter (CNM) effects and include  parton energy loss and multiple scattering before the hard scattering, and modifications of the parton distribution functions in the nucleus (nPDFs) with respect to those of a free nucleon (PDFs). Some nPDF modifications have been previously deduced from measurements of lepton-nucleus deep inelastic scattering and Drell--Yan production of lepton pairs from $\qqbar$ annihilation in proton-nucleus collisions~\cite{Arneodo1994301}. In addition, measurements of $\pi^{0}$ production in deuteron-gold collisions at RHIC~\cite{Adler:2006wg} are also included in recent nPDF fits to better constrain the nuclear gluon distributions~\cite{Eskola:2009uj}. There are several ranges in the parton fractional momenta $x$ in which the data show suppression or enhancement in the nPDFs relative to the proton PDFs. At small $x$ (${\lesssim}0.01$), the nPDFs are found to be suppressed, a phenomenon commonly referred to as ``shadowing''~\cite{Frankfurt:1988nt}. In the range $0.02 \lesssim x \lesssim 0.2$, the nPDFs are enhanced (``antishadowing''~\cite{Arneodo1994301}), and for $x \gtrsim 0.2$ a suppression  has been seen (``EMC effect"~\cite{Norton:2003cb}).

Proton-lead (\pPb) collisions at the LHC provide an opportunity to evaluate the CNM effects and establish an additional reference for the interpretation of  measurements performed in \PbPb  collisions. The results of several \pPb studies involving jets or dijets~\cite{ATLAS:2014cpa,Chatrchyan:2014hqa,Adam:2015xea}, electroweak bosons~\cite{Aad:2015gta,Khachatryan:2015hha}, and high \pt charged particles~\cite{Khachatryan:2015xaa, Abelev:2014dsa} are already available. No significant indication of jet quenching was found so far in the \pPb studies of inclusive jet production~\cite{ATLAS:2014cpa,Adam:2015hoa},  dijet momentum balance~\cite{Chatrchyan:2014hqa}, dijet acoplanarity~\cite{Chatrchyan:2014hqa, Adam:2015xea},  or charged-hadron measurements~\cite{Khachatryan:2015xaa, Abelev:2014dsa}. The shapes of the  dijet~\cite{Chatrchyan:2014hqa}  and Z boson~\cite{Aad:2015gta}  pseudorapidity distributions are found to be in better agreement with EPS09 nPDF  predictions~\cite{Eskola:2009uj} than with the free-proton PDFs for measurements inclusive in the  impact parameter. Hints of modifications  larger than those presently included in the EPS09 nPDFs have also been seen~\cite{Aad:2015gta,Khachatryan:2015hha,Khachatryan:2015xaa}. In particular, the charged hadron spectra~\cite{Khachatryan:2015xaa} are found to be enhanced at high \pt~beyond the anti-shadowing included in  EPS09.  Significant modifications with respect to those included in EPS09 have also been found for  impact-parameter-dependent measurements~\cite{Chatrchyan:2014hqa,ATLAS:2014cpa}. The interpretation of the latter results is more difficult because of the kinematic biases introduced through the event selections~\cite{Chatrchyan:2014hqa,ATLAS:2014cpa,Armesto:2015kwa,Adare:2013nff,Adam:2014qja}.

In this paper we present the CMS measurements of inclusive jet production in \pPb collisions at a nucleon-nucleon (NN) center-of-mass energy of $\sqrtsNN=5.02$\TeV as a function of \pt in several pseudorapidity regions in  the range  $-2.0 <\eta_\mathrm{CM}<1.5$ in the NN center-of-mass system. No additional event activity  selections have been made to avoid the associated kinematic biases. The measurements extend in  \pt up to 500\GeVc and are sensitive  to nPDF modifications in the anti-shadowing and EMC effect regions. Since presently there are no experimental results available from \pp collisions at $\sqrt{s} = 5.02$\TeV, \pp reference jet spectra in pseudorapidity ranges corresponding  to the present measurements are obtained by extrapolating jet measurements at  $\sqrt{s} = 7$\TeV~\cite{Chatrchyan:2014gia}. The paper is organized as follows: Section~\ref{sec:analysis} provides the experimental details, Section~\ref{sec:sys} gives an account of the systematic uncertainties in the measurements, Section~\ref{sec:results} presents the results, and Section~\ref{sec:summary} summarizes our findings.

\section{Data analysis}
\label{sec:analysis}
This measurement is based on a data sample of \pPb collisions corresponding to an integrated luminosity of 30.1\nbinv collected by the CMS experiment in 2013.
The beam energies were 4\TeV for protons and 1.58\TeV per nucleon for lead nuclei, resulting in a center-of-mass
energy per nucleon pair of 5.02\TeV. The direction of the higher-energy proton
beam was initially set up to be clockwise within CMS conventions, and was reversed after a data set corresponding to an integrated luminosity of
21\nbinv  was recorded.
 As a result of the energy difference of the colliding beams, the nucleon-nucleon center-of-mass in the \pPb collisions is shifted with respect to zero rapidity in the laboratory frame.
 Both portions of the data set are analyzed independently and the results are found to be compatible
within their uncertainties.
In order to reduce the statistical uncertainties, the two data sets are then combined. Results from the first data taking period are reflected along the $z$-axis so that in the combined analysis the proton travels in the positive $z$ and pseudorapidity $\eta$ direction. In the laboratory frame $\eta = -\ln[\tan(\theta/2)]$, where $\theta$ is the polar angle defined with respect to the proton beam direction. The results are presented in this convention, after transformation to the NN center-of-mass frame, which for massless particles is equivalent to a shift in pseudorapidity: $\eta_\mathrm{CM} = \eta - 0.465$.

\subsection{Experimental setup}
\label{sec:experiment}

A detailed description of the CMS detector and of its coordinate system can be found in  Ref.~\cite{CMS_Detector}.
It features nearly hermetic calorimetric coverage and high-resolution tracking
for the reconstruction of energetic jets and charged particles.
The calorimeters consist of a lead tungstate crystal electromagnetic calorimeter (ECAL), and a brass and scintillator hadron
calorimeter (HCAL) with coverage up to $\abs{\eta} = 3$.
The quartz/steel hadron forward (HF) calorimeters extend the calorimetry coverage in the region $3.0< \abs{\eta}< 5.2$, and are used in offline event selection.
The calorimeter cells are grouped in projective towers of granularity
$\Delta\eta{\times}\Delta\phi = 0.087{\times}0.087$ (where $\phi$ is the azimuthal angle in radians) for the central
pseudorapidity region used in the present  jet measurement, and have coarser segmentation (about twice as large) at forward pseudorapidity.
The central calorimeters are enclosed in a superconducting solenoid with 3.8\unit{T} magnetic field.
Charged particles are reconstructed by the tracking system, located inside the calorimeters and the superconducting coil. It consists of
silicon pixel and strip layers covering the range $\abs{\eta}< $ 2.5,  and provides track reconstruction with
momentum resolution of about 1.5\%  for high-\pt particles.  Muons are measured in gas-ionization detectors embedded in the steel flux-return yoke outside the solenoid.

\subsection{Event selection}

The CMS online event selection employs a hardware-based level-1 (L1) trigger and a software-based high-level trigger (HLT).
A minimum bias sample is selected by the L1 requirement
of a pPb bunch crossing at the interaction point and an HLT requirement of at least one reconstructed track with
$\pt > 0.4$\GeVc in the pixel tracker. This minimum bias trigger was prescaled by a large factor for most of the 5.02\TeV data collection, because of the high instantaneous luminosity of the LHC.
In order to increase the \pt range of the measurement, additional HLT triggers were
used to select events based on the presence of a jet with $\pt > 20$, 40, 60, 80, or 100\GeVc reconstructed in the calorimeters.

\begin{figure}[htbp!]
\centering
\includegraphics[width=.48\textwidth]{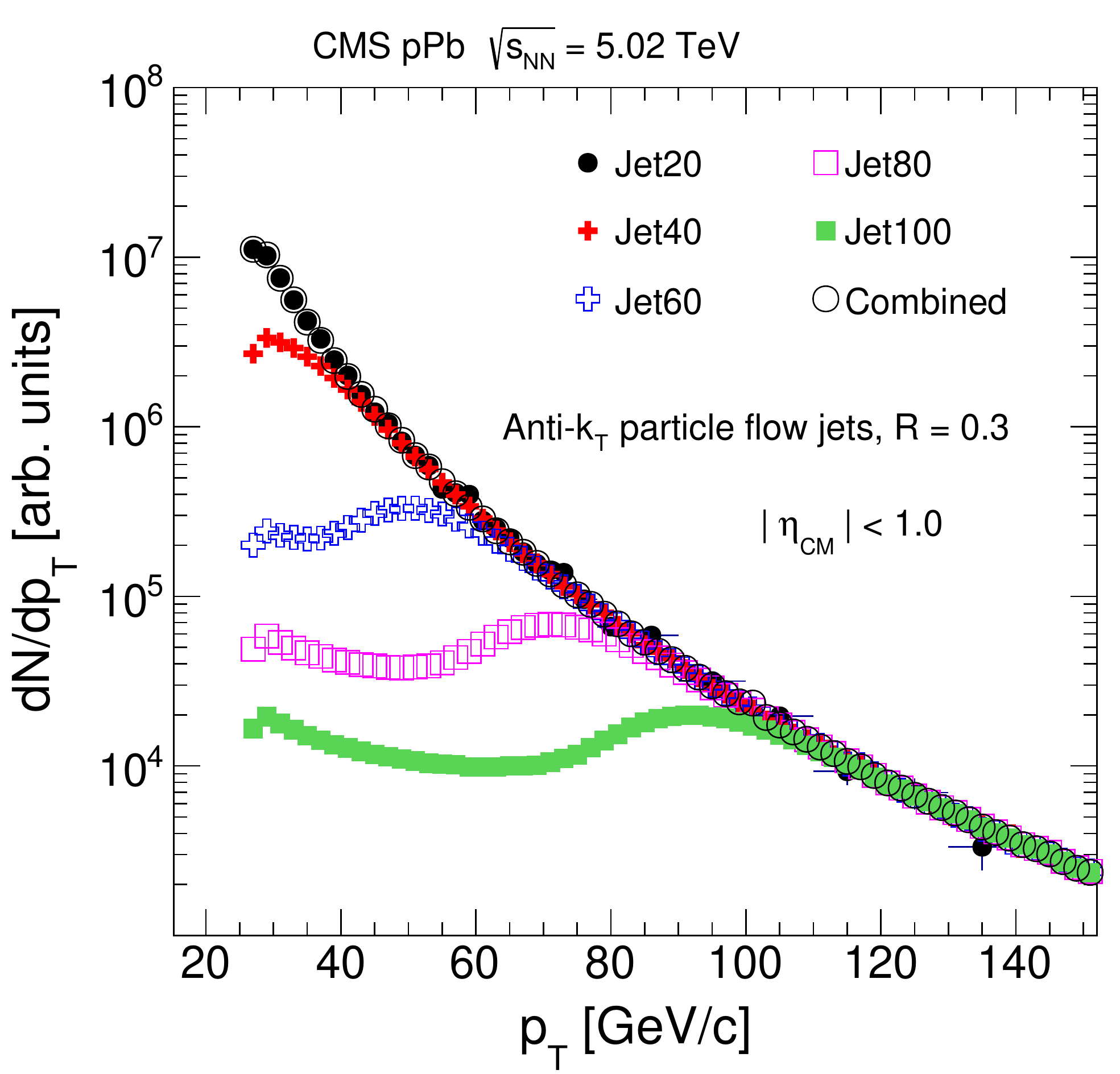}
\includegraphics[width=.48\textwidth]{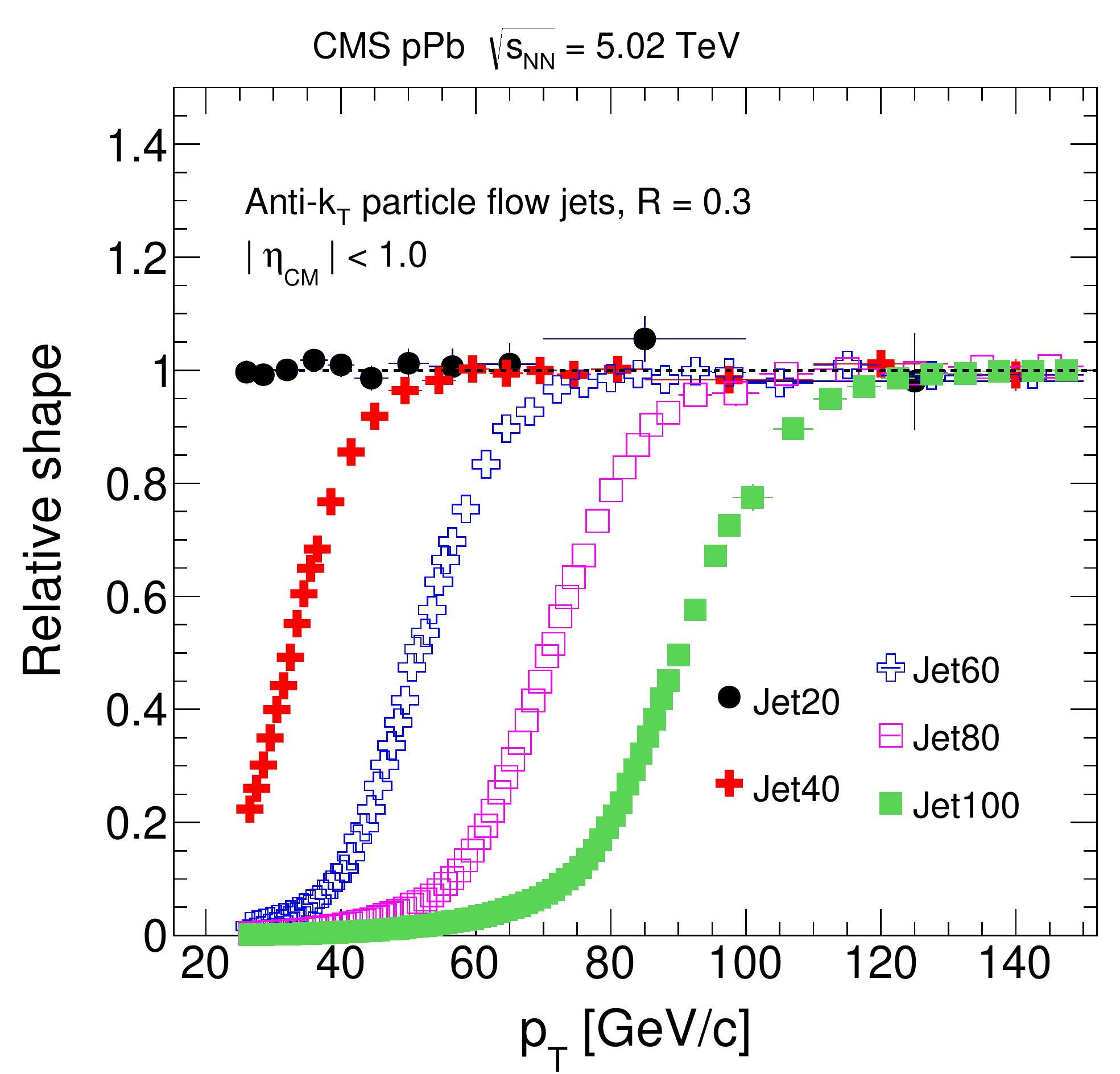}
\caption{\cmsLeft: The weighted jet spectra using prescale factors from each HLT-triggered event sample and the combined jet spectrum. A subset of the data is plotted to illustrate the procedure.
\cmsRight: The ratios of each individual HLT-triggered jet spectrum to the combined jet spectrum. Statistical uncertainties are shown as  vertical bars, and \pt bin widths as horizontal bars.}
\label{fig:TrigComb}
\end{figure}

For the offline analysis, an additional selection of hadronic collisions
is applied by requiring a coincidence of at least one HF calorimeter
tower with more than 3\GeV of total energy on the positive and negative sides of the interaction point. Events are further required to have at least one reconstructed primary
vertex with at least two associated tracks~\cite{Khachatryan:2010xs}.  A maximum distance of 15\unit{cm} between the primary vertex and the nominal interaction point along the beam line is required to ensure maximum tracking acceptance. Additionally,  track-based selection cuts are applied to
suppress of beam-related background events~\cite{Khachatryan:2010pw}.
The instantaneous luminosity of the pPb run in 2013 resulted in a 3\% probability of at least one additional interaction occurring in the same bunch crossing. Events with more than one interaction (``pileup" events) are removed using  a rejection algorithm developed in  Ref.~\cite{Khachatryan:2015xaa}. The pileup-rejection efficiency of this filter is found to be  $90\pm2$\% in minimum bias events and it removes a very small fraction (0.01\%) of the events without pileup.
In order to combine the spectra measured from the various jet-triggered data samples, the events included
in the analysis are weighted according to the individual HLT prescale factors corresponding to the trigger object
with maximum \pt in the event. The \cmsleft  panel of Fig.~\ref{fig:TrigComb} shows the prescale-weighted jet spectra that are reconstructed with the anti-\kt~\cite{Cacciari:2008gp} algorithm from each HLT trigger path and the combined inclusive jet spectrum. The ratios of each HLT-triggered spectrum to the combined jet spectrum
are shown in the \cmsright   panel of Fig.~\ref{fig:TrigComb}. In the range of \pt where the triggers are fully efficient, this ratio is unity and independent of jet \pt.

\subsection{Jet reconstruction and corrections}
\label{sec:jets}

The CMS particle-flow (PF) algorithm~\cite{particle_flow,CMS-PAS-PFT-10-002} identifies
 stable particles in an event by combining information from all sub-detector systems,  classifying them as
electrons, muons, photons, and charged and neutral hadrons. The PF candidates
are then clustered into jets using the anti-\kt sequential recombination algorithm~\cite{Cacciari:2008gp} provided in the \textsc{FastJet}
framework~\cite{Cacciari:fastjet1}. The results in this analysis are obtained using a distance parameter $R=0.3$.
The underlying event (UE) contribution to the jet energy is subtracted using an
iterative procedure described in Refs.~\cite{Kodolova:2007hd,Chatrchyan:2011sx}. The jet energies are then corrected to contain the energy of all final-state jet constituents as described in Ref.~\cite{Khachatryan:2015luy}. The jet energy corrections are derived using simulated  \PYTHIA (6.462, Z2 tune)~\cite{Sjostrand:2006za,Field:2011iq} events and measurements of the energy balance of dijet and photon+jet \pPb collision events are used to correct differences between data and Monte Carlo (MC) distributions~\cite{Chatrchyan:2014hqa,Khachatryan:2015luy}.
In the jet reconstruction process, there is a possibility that the jet energy is estimated incorrectly, or a jet is found in a region where the UE has an upward fluctuation, but no hard scattering has occurred (a ``fake'' jet). In MC the ``real'' and ``fake'' jets can be distinguished by requiring that the reconstructed jet is matched to a generator-level jet. In data, this cannot be done directly, but the contribution of fake jets could be estimated from  MC, provided that it is tuned to describe the data, and specific jet selections are developed to identify and remove the  misreconstructed jets.  We estimate that about 10\% of the jets reconstructed at \pt = 50 \GeVc in \pPb collisions are fake, and this fraction quickly drops to a level of $10^{-4}$ at \pt$\approx~100$ \GeVc. After the jet-identification cuts are applied, we estimate that less than 1\% fake jets remain in the sample.

\begin{figure}[h!]
\centering
\includegraphics[width=.48\textwidth]{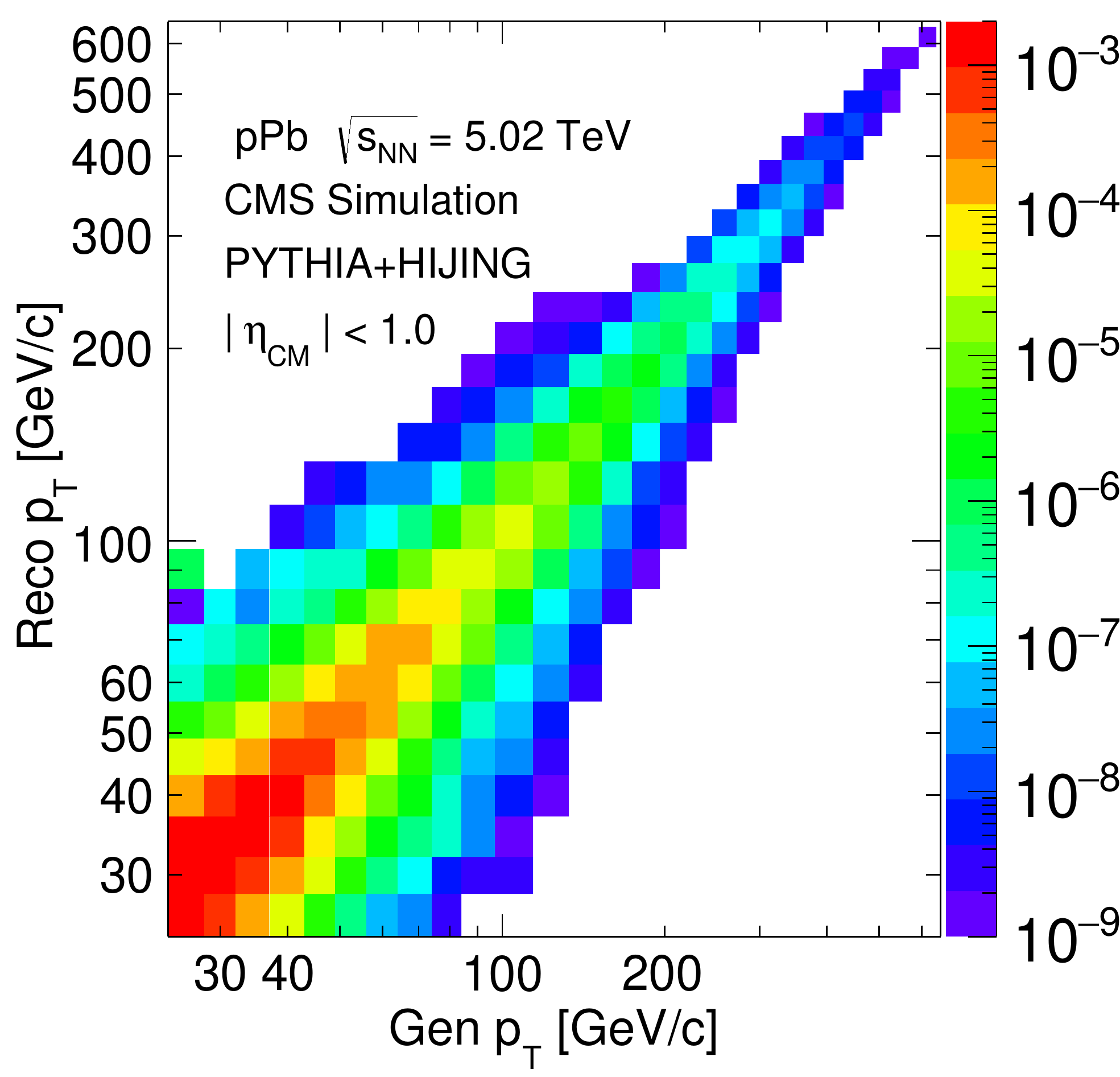}
\includegraphics[width=.48\textwidth]{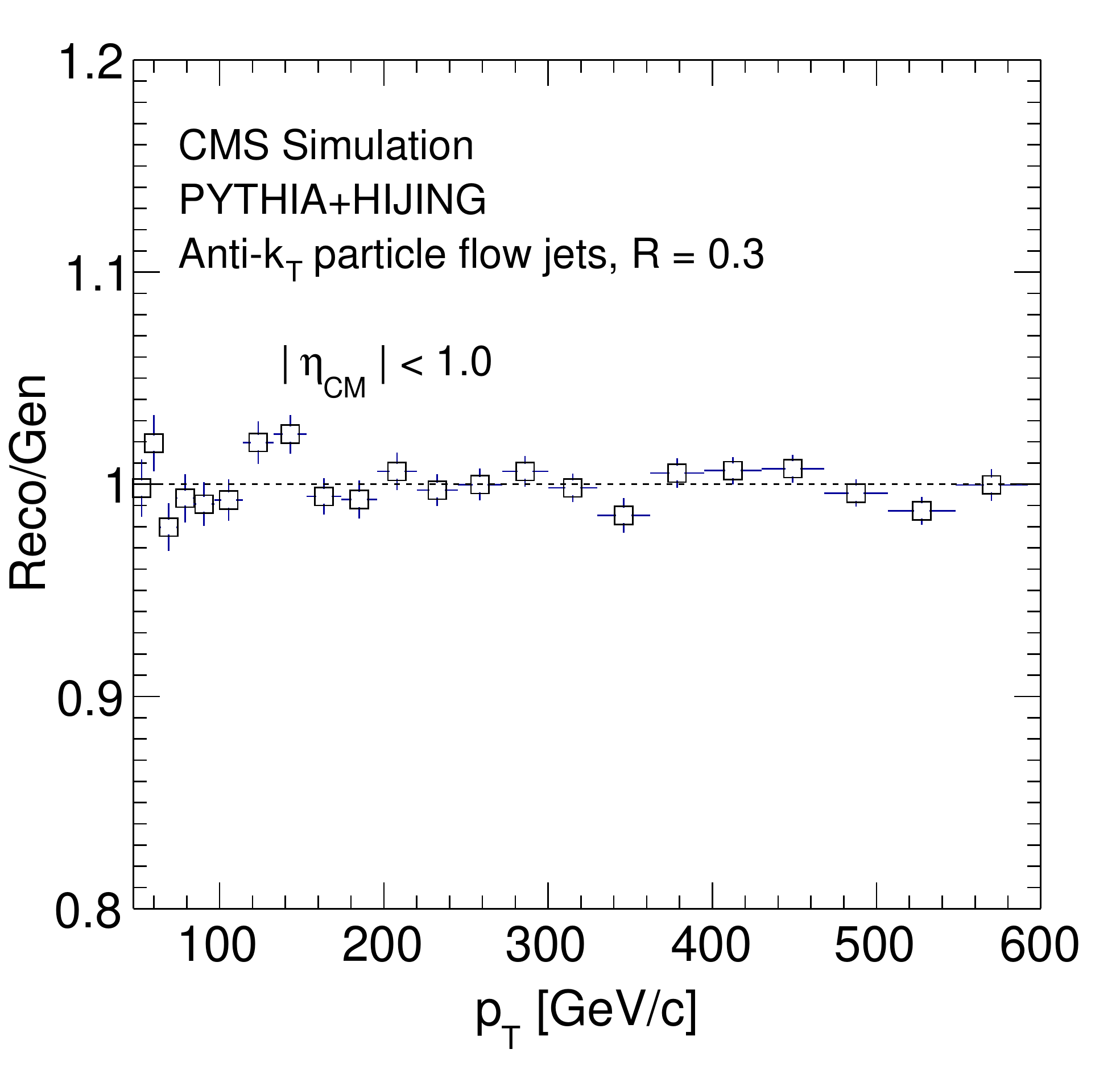}
\caption{
\cmsLeft: Response matrix built from {\PYTHIAHIJING} simulation. \cmsRight:  The
ratios of the Bayesian unfolded jet \pt spectrum reconstructed in the simulation
and the generator-level spectrum.}
\label{fig:MatrixClosure}
\end{figure}

Because of the finite detector resolution and the steeply falling \pt distributions, the measured  jet \pt spectra are smeared with respect to the true distributions, although the mean value of the reconstructed jet energy is corrected as described above.  The jet energy resolution is estimated to be 13\%\,(8\%) for jet $\pt = 60$\,(300)\GeVc.
A Bayesian unfolding technique~\cite{D'Agostini:1994zf} is employed to account for such resolution effects, as implemented in the RooUnfold package~\cite{Adye:2011gm}. The migration of jets in pseudorapidity is not explicitly corrected for; it is instead included as an uncertainty, as discussed in Section~\ref{sec:syspPb}.
In the unfolding method, a response matrix is built based on MC simulations and is used to obtain the
``true''  jet \pt distribution from the measured one. Jets are first generated with the \PYTHIA
event generator and then embedded into \pPb collisions simulated with the {\HIJING} event generator (version 1.383)~\cite{hijing}, which have particle multiplicity distributions comparable to the \pPb data and can account for additional resolution effects associated with the higher detector occupancy. These embedded MC samples are denoted hereafter by {\PYTHIAHIJING}.
The unfolding technique is tested  by building the response matrix with detector jets (Reco) and generated jets (Gen) from half of the MC sample and applying it to unfold the other half of the
sample. The
\cmsleft~ panel  of Fig.~\ref{fig:MatrixClosure} shows the response matrix obtained using the {\PYTHIAHIJING} simulation, while the \cmsright  panel shows the ratio of the jet spectrum reconstructed from the simulation after unfolding to the generator-level jet spectrum.
The unfolded MC jet spectrum is compatible with the generator-level jet spectrum within the statistical uncertainties.
The results reported in this paper are based on the Bayesian unfolding technique that uses four iteration steps. Up to eight iteration steps are used in evaluating the systematic uncertainties as discussed in Section~\ref{sec:syspPb}. The generator level \PYTHIA jet spectrum is used as a prior in the unfolding.
The  data points are reported in the center of each  \pt bin without  corrections for binning effects.

The \pPb jet cross sections are obtained in several pseudorapidity intervals. To study the evolution of the jet cross section with pseudorapidity, ratios of jet spectra are computed either using symmetric positive and negative pseudorapidity intervals around mid-rapidity, or normalizing the distributions by the mid-rapidity jet spectrum. These ratios are taken in the same  \pt bin and the values are reported at the center of the bin. To study nuclear effects on jet production, the jet spectra in \pPb collisions are compared to \pp reference spectra obtained by extrapolation from previous jet cross section measurements in \pp collisions  at higher center-of-mass energy. The nuclear modification factor, $R_\pPb$, evaluated in several pseudorapidity intervals, is defined as

\begin{equation}
R_\pPb=\frac{1}{A}\frac{{\rd}^{2}\sigma_\text{jet}^\pPb/{\rd} \pt\,{\rd} \eta}{{\rd}^{2}\sigma_\text{jet}^{\pp}/{\rd} \pt\,{\rd}\eta} =\frac{1}{A}\frac{1}{L}\frac{{\rd}^{2}N_\text{jet}^\pPb/{\rd}\pt\,{\rd}\eta}{{\rd}^{2}\sigma_\text{jet}^{\pp}/{\rd}\pt\,{\rd}\eta},
\end{equation}

where  $L=30.1$\nbinv is the effective integrated luminosity in the \pPb analysis, corrected for event-selection efficiency and trigger prescales, and $A$ is the mass number of the lead nucleus. Since presently there are no available experimental results from \pp collisions at  $\sqrt{s} = 5.02$\TeV,  for this paper we use extrapolated, rather than measured, \pp reference spectra. Hence we denote the nuclear modification factors  as $R^{\ast}_\pPb$.

\subsection{Proton-proton reference jet spectra}

\begin{figure}[h!tb]
\centering
\includegraphics[width=\cmsFigWidth]{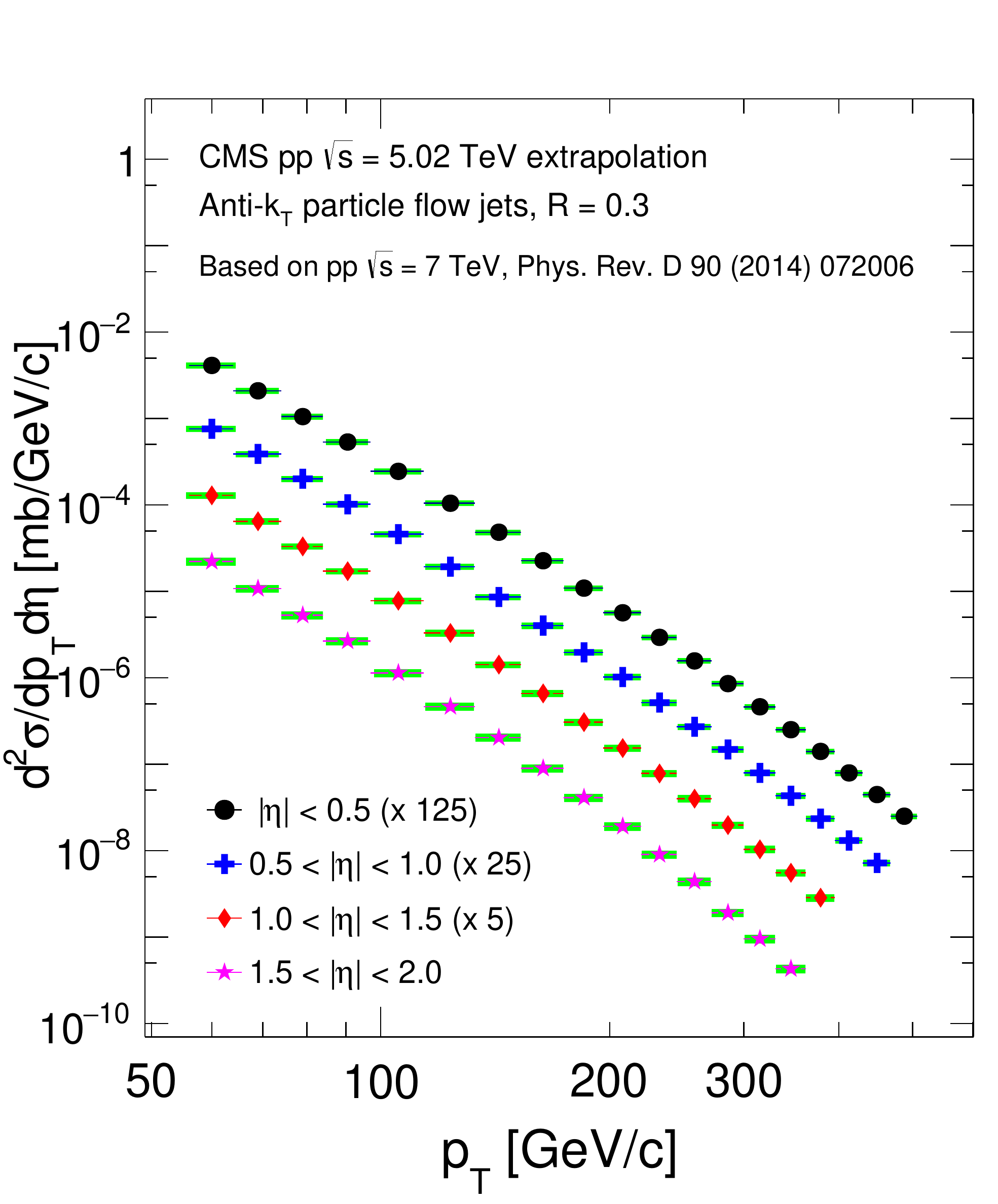}
\caption{
Jet spectra at $\sqrt{s} = 5.02$\TeV extrapolated from previous \pp measurements at $\sqrt{s} = 7$\TeV~\cite{Chatrchyan:2014gia}. Additional scaling factors listed in the legend are  applied to enhance the visibility. The horizontal bars represent the bin size, and the points are plotted in the center of the bin. The shaded boxes denote the systematic uncertainties in the extrapolation procedure. The statistical  uncertainties are smaller than the symbol size. }
\label{fig:ppReference}
\end{figure}

The reference \pp spectra are constructed extrapolating previously published inclusive jet spectra measured in \pp collisions at
$\sqrt{s}=7$\TeV. Measurements  performed with the anti-\kt jet algorithm with two distance parameters, $R=0.5$
and $0.7$~\cite{Chatrchyan:2014gia}, are used in the extrapolation. The extrapolation is based on the \PYTHIA generator (6.462, Z2 tune) and is performed in two steps. First, the $\sqrt{s}=7$\TeV jet cross section measurements are  extrapolated
to $\sqrt{s} = 5.02$\TeV and then scaled to $R = 0.3$, since a smaller distance parameter is used in the \pPb analysis
to minimize the UE background fluctuations. The \PYTHIA generator is used to estimate \pt -dependent scaling factors.
While this scaling is model dependent, the ratio of the jet cross sections measured with $R=0.5$ and $0.7$ appears to be well reproduced in \PYTHIA within 3\%~\cite{Chatrchyan:2014gia}. Several alternative methods are used to derive cross section scaling factors in $\sqrt{s}$ and in distance parameter in order to evaluate the systematic uncertainties discussed in Section~\ref{sec:syspp}. The extrapolated jet spectra are shown in Fig.~\ref{fig:ppReference}. Scaling factors are applied, as noted in the legend, to enhance the visibility.

\section{Systematic uncertainties}
\label{sec:sys}
\subsection{Systematic uncertainties in the \pPb measurement}
\label{sec:syspPb}

There are several sources of systematic uncertainty in the measurements of the jet spectra, the jet yield asymmetry, and
the nuclear modification factors $R^{\ast}_\pPb$.  The dominant uncertainties in the spectra measured in \pPb collisions come from the unfolding of the spectra and from the jet energy scale (JES) corrections, which are partially correlated since they both aim to correct  for the difference between the reconstructed and the true jet energy. The stability of the unfolding procedure and its ability to recover the generator-level jet spectrum have been verified with simulation studies, which included the use of different numbers of iterations ($n=2$, 3, 4, 5, 6, 7, 8). In the data, the unfolded spectra for $n=4$ are compared to the spectra obtained with different values of $n$ and the difference is included in the systematic uncertainty. In addition, since the true jet spectrum may differ in shape from the spectrum in the MC generator,  the slope of the prior guess distribution is varied such that the yield at low \pt increases or decreases by a factor of 3, while at high \pt the yield is changed by about 10--20\%.  After this variation the spectra are unfolded and then are compared to the nominal unfolded spectra to estimate the uncertainty due to the nominal input distribution. The uncertainties from unfolding are largest (up to 5\%) in the low \pt region and at large absolute pseudorapidity.  Uncertainties that arise from the different jet energy resolution in the data and MC simulation are evaluated by smearing the unfolding matrix to account for these differences and then redoing the unfolding. The resulting differences in the final jet spectra are found to be less than 1\%. The JES uncertainty is about 1\% and induces up to 7\% changes at high \pt because of the steeply falling jet spectra.

Additional cross checks are performed comparing the spectra obtained with different  jet reconstruction algorithms (such as subtracting the UE  in the jet algorithm or correcting for it in the transfer matrix), and comparing the unfolded results when the unfolding matrix uses the reconstructed jet \pt with or without jet energy corrections. The total uncertainty in the jet spectra due to the JES and unfolding varies from about 5\% at low jet \pt at mid-rapidity to about 10\% for high \pt and forward rapidity.

The fake jet contribution is estimated on the basis of a MC study of various jet quality variables
that are used to identify genuine and misreconstructed jet contributions. In the {\PYTHIAHIJING} embedded samples these variables are optimized to
remove misreconstructed jets, while preserving the largest fraction of genuine jets. The uncertainty in the misreconstructed jet contribution in the jet spectra is estimated by varying the jet quality requirements and comparing the resulting spectra in data and in simulation. It is about 1\% for all pseudorapidity ranges.

The unfolding procedure does not correct for possible misreconstruction of the jet axis, and therefore jets may migrate from one pseudorapidity interval to another thus altering the jet spectra measured in different $\eta$ ranges. The uncertainty associated with the  jet pointing resolution  is estimated by building the unfolding matrix using either the generated or the reconstructed jet axis, and comparing the resulting unfolded jet spectra. This uncertainty is found to be of the order of 1\% in the central pseudorapidity region and 2\% at large absolute pseudorapidity.

The jet spectra in \pPb collisions are also subject to an overall scale uncertainty, due to the uncertainties in the integrated luminosity measurement. The scale uncertainty is estimated to be  3.5\%, as described in Ref.~\cite{Khachatryan:2015zaa}.

The systematic uncertainty in the inclusive jet production asymmetry only includes those
factors that depend on the jet pseudorapidity,  such as the JES, unfolding,  and misreconstructed jet contribution uncertainties.
The overall scale uncertainty due to the luminosity normalization cancels out. As a cross check, the jet yield asymmetry uncertainties
are evaluated using a combination of the two data sets with different beam directions. In that case, the jet yield asymmetry can be measured  using detector elements that are only in the positive $\eta$ or in the negative $\eta$ ranges in the laboratory frame. Since the detector is symmetric, these regions have similar acceptance and performance and we expect that systematic effects are also similar. Alternatively, the jet yield asymmetry is measured from each portion of the data independently, and the results of this comparison confirm the systematic uncertainty estimate obtained by evaluating each source of uncertainty separately.

\subsection{Systematic uncertainties in the \pp reference}
\label{sec:syspp}

The uncertainties in the extrapolated \pp reference spectra take into account the uncertainties in the distance parameter dependence of the cross sections at  $\sqrt{s}= 7 $\TeV and the scaling to the smaller $R=0.3$ value, the uncertainty in the  $\sqrt{s}$ dependent scaling, as well as the uncertainties of the input spectra used in the extrapolation. The uncertainties in the inclusive jet measurements from \pp collisions at \mbox{$\sqrt{s} = 7$\TeV} reported in Ref.~\cite{Chatrchyan:2014gia} are taken as the upper and lower limits of the cross sections used in the extrapolation, and are reflected in the uncertainties of the resulting reference spectra. The following alternative approaches are used to derive scaling factors and evaluate their uncertainties.
\begin{enumerate}
\item{\PYTHIA 8, CUETP8M1 tune~\cite{Sjostrand:2007gs,Khachatryan:2015pea}

 In the kinematic range studied, this tune has a different quark-to-gluon jet ratio and different jet shapes than the \PYTHIA 6, Z2 tune used for the nominal result. }
\item { {\POWHEGPYTHIA} event generator~\cite{Alioli:2010xd,Alioli:2010xa}

The \POWHEG generator is used to compute the cross section at next-to-leading order (NLO) accuracy, and \PYTHIA (6.462, Z2 tune) is used to  describe the parton showering and  hadronization.}
\item{ NLO calculations~\cite{Nagy:2001fj, Nagy:2003tz} with several different parametrizations of the parton distribution functions~\cite{Britzger:2012bs} and non-perturbative corrections based on \PYTHIA (6.462, Z2 tune). }
\item{ Jet cross section measurements with  $R=0.7$ at  $\sqrt{s}=7$\TeV~\cite{Chatrchyan:2014gia} and  $\sqrt{s}=2.76$\TeV~\cite{Khachatryan:2015pp} are used to evaluate $\sqrt{s}$ dependent scaling factors using  $x_\mathrm{T}$-based interpolation ($x_\mathrm{T}\equiv 2p_\mathrm{T}c/\sqrt{s}$).}
\end{enumerate}

The jet cross sections for  $R=0.3$ and  $R=0.5$  at  $\sqrt{s}= 5.02$ TeV are  evaluated using (1), (2), and (3). Then the ratios between the cross sections obtained with these two distance parameters, in the default \PYTHIA calculation (6.462, Z2 tune) and in the alternative methods, are compared to each other, leading to an uncertainty in the distance parameter scaling of around 5\%. The $\sqrt{s}$ scaling factors are evaluated with (2) and (3) for $R=0.5$,  and with (2), (3), and (4) for $R=0.7$. These scaling factors are compared to the results from \PYTHIA (6.462, Z2 tune). The uncertainties in the $\sqrt{s}$ scaling factors range from 4\% at low jet \pt in the mid-rapidity region to 7\% at high \pt and at forward rapidity. The total uncertainty in the pp reference extrapolation is found to range between 9\% at mid-rapidity and 11\% at forward rapidity.
These uncertainties include a 2.4\% scale uncertainty from the integrated luminosity measurement~\cite{Chatrchyan:2014gia}.

\subsection{Summary of systematic uncertainties}
\label{sec:sysTotal}

A summary of the systematic uncertainties in the jet spectra in \pPb collisions, the jet yield asymmetry measurements in \pPb collisions, the reference \pp spectra, and the nuclear modification factors $R^{\ast}_\pPb$  are listed in Table~\ref{tab:sys}.  The uncertainties depend on the jet \pt and pseudorapidity, and the table shows representative values in two jet \pt and $\eta_\mathrm{CM}$ ranges. The uncertainties vary smoothly between these ranges. The total systematic uncertainties listed for the nuclear modification factors $R^{\ast}_\pPb$ do not include the scale uncertainty of 4.3\%  from the integrated luminosity measurements in \pPb (3.5\%) and \pp (2.4\%) collisions. The luminosity uncertainties cancel in the measurements of the jet yield asymmetry.
The remaining uncertainties are partially correlated in jet \pt, with the unfolding uncertainty dominating at low jet \pt and the JES uncertainty dominating at high jet \pt.

\begin{table*}[htbp]
\centering
\topcaption{Systematic uncertainties in the measurement of the jet spectra in \pPb collisions are shown in the first four lines. The sources and corresponding systematic uncertainties in the extrapolated \pp reference are presented in the next four lines. The total uncertainties in the jet spectra in \pPb collisions, the reference \pp spectra, the jet yield asymmetry in \pPb collisions, and $R^{\ast}_\pPb$ are shown in the bottom four lines. The uncertainties depend on the jet \pt and pseudorapidity, and the table shows representative values in two jet \pt and $\eta_\mathrm{CM}$ ranges. The uncertainties vary smoothly between these two ranges. Total systematic uncertainties listed for the nuclear modification factors $R^{\ast}_\pPb$ do not include the scale uncertainty of 4.3\% due to the uncertainty in the integrated luminosity measurements in \pPb (3.5\%) and \pp (2.4\%) collisions.}
\label{tab:sys}
\begin{tabular}{llrr{c}@{\hspace*{5pt}}rr}
&\multirow{2}{*} {Source}
&\multicolumn{2}{c} {Jet $\pt < 80$\GeVc } && \multicolumn{2}{c} {Jet $\pt > 150$\GeVc} \\ \cline{3-4}\cline{6-7}
&& $\abs{\eta_\mathrm{CM}}< 1$  & $\abs{\eta_\mathrm{CM}}> 1.5$ && $\abs{\eta_\mathrm{CM}}< 1$  & $\abs{\eta_\mathrm{CM}}> 1.5$\\
\hline
\pPb: &JES \& unfolding     & 5\%  & 8\% &&7\%  & 10\%\\
&Misreconstructed jet contribution& 1\% & 1\% &&1\%  & 1\%\\
&Jet pointing resolution    & 1\%  & 2\% &&1\%  & 2\%\\
 &Integrated luminosity     & 3.5\%  & 3.5\% &&3.5\% &  3.5\%  \\
\hline
\pp: &Input data            & 6\%  & 8\% &&5\%  & 7\%\\
&Cone-size dependence       & 5\%  & 5\% &&5\%  & 5\%\\
&Collision-energy dependence& 4\%  & 5\% &&6\%  & 7\%\\
 &Integrated luminosity     & 2.4\%& 2.4\% &&2.4\% &2.4\%  \\
\hline
Total: & \pPb spectra      & 6\%  & 9\% &&8\%   & 11\%\\
& \pPb asymmetry           & 7\%  & 11\% &&10\% & 14\%\\
& \pp reference            & 9\%  & 11\% &&10\%  & 11\%\\
&  $R^{\ast}_\pPb$          & 10\%  & 14\% &&12\% & 15\%\\
\end{tabular}
\end{table*}

\section{Results and discussion}
\label{sec:results}
\begin{figure}[hpbt]
\centering
\includegraphics[width=0.5\textwidth]{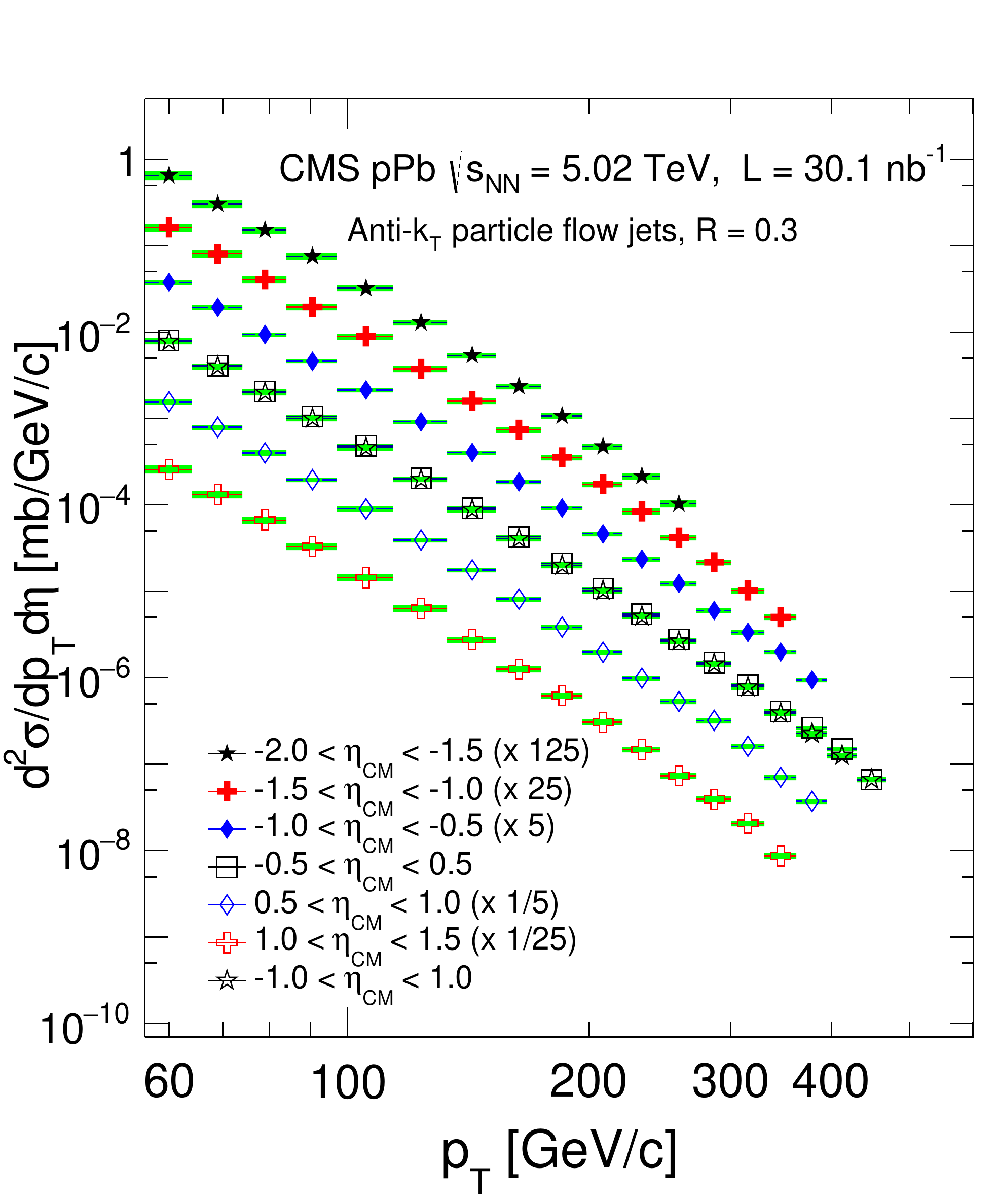}
\caption{\label{fig:UnfoldSpectra}
Inclusive jet differential cross section in \pPb collisions at $\sqrtsNN=5.02$\TeV in six consecutive eta bins plus the range $\abs{\eta_\mathrm{CM}} < 1.0$. The spectra are scaled by arbitrary factors for better visibility.  The horizontal bars represent the bin width, and the filled boxes indicate the systematic uncertainties. The statistical uncertainties are  smaller than the symbol size.  }
\end{figure}

\begin{figure}[hpbt]
\centering
    \includegraphics[width=0.5\textwidth]{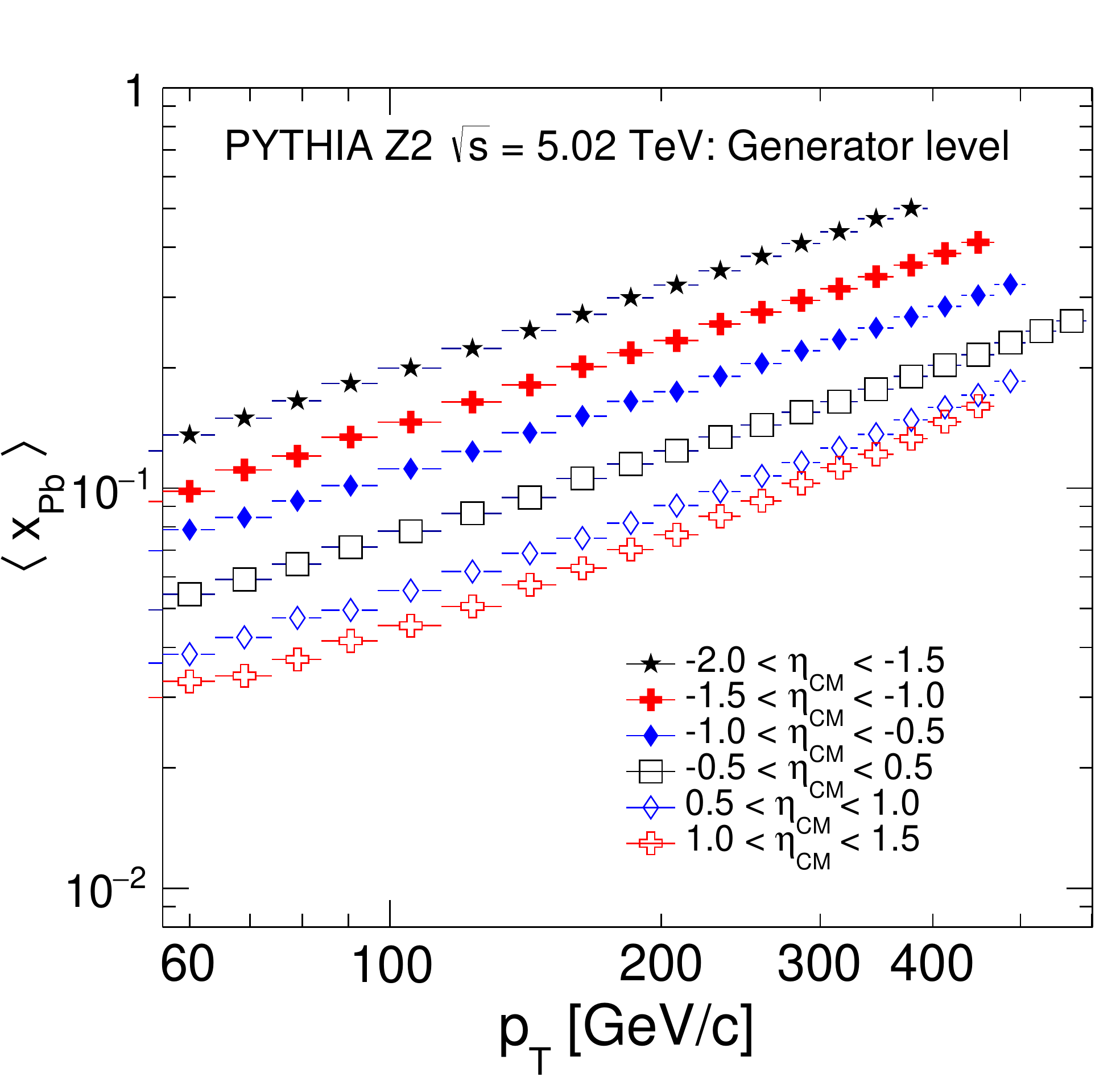}
\caption{\label{fig:xpb}
Mean $x$ values of partons in the Pb nucleus, $\langle x_\Pb\rangle$, corresponding to the jet \pt and pseudorapidity ranges covered in the measurements.  The   $\langle x_{\Pb}\rangle$ values are determined using the \PYTHIA event generator~\cite{Sjostrand:2006za}.  }
\end{figure}
The inclusive jet differential cross sections in \pPb
collisions at $\sqrtsNN=5.02$\TeV are shown in
Fig.~\ref{fig:UnfoldSpectra} for six consecutive $\eta$ intervals in the range  $-2.0 <\eta_\mathrm{CM}<1.5$ and the range $\abs{\eta_\mathrm{CM}}<1.0$ for reference purposes. The distributions are scaled by arbitrary factors described in the legend to enhance visibility. These spectra are used to study the pseudorapidity dependence of inclusive jet production in \pPb collisions and possible nuclear effects. In symmetric collisions, such as in the \pp system, the kinematic range in the fractional momentum  $x$ probed with the jets in forward and backward pseudorapidity is the same and the production is symmetric about $\eta_\mathrm{CM}=0$. In the \pPb system, the jets produced at forward pseudorapidity (proton beam direction)  correlate with smaller $x$ values from the Pb nucleus than those produced at backward pseudorapidity.
Based on a generator-level study made with \PYTHIA, the average $x$ values from the Pb nucleus (Fig.~\ref{fig:xpb}) that are probed in the kinematic range covered by the present measurement  are estimated to be in the range $0.03\lesssim \langle x_\Pb\rangle \lesssim 0.5$. Values of \pt that correspond to $\langle x_\Pb\rangle \lesssim 0.2$  are associated with anti-shadowing in the nPDFs. The region $\langle x_\Pb\rangle \gtrsim 0.2$ is associated with a suppression in the nPDFs with respect to the free-nucleon PDFs (EMC effect), and can be reached at high jet \pt in the backward pseudorapidity region ($\eta <-1$).

\begin{figure}[hpbt]
\centering
    	\includegraphics[width=0.5\textwidth]{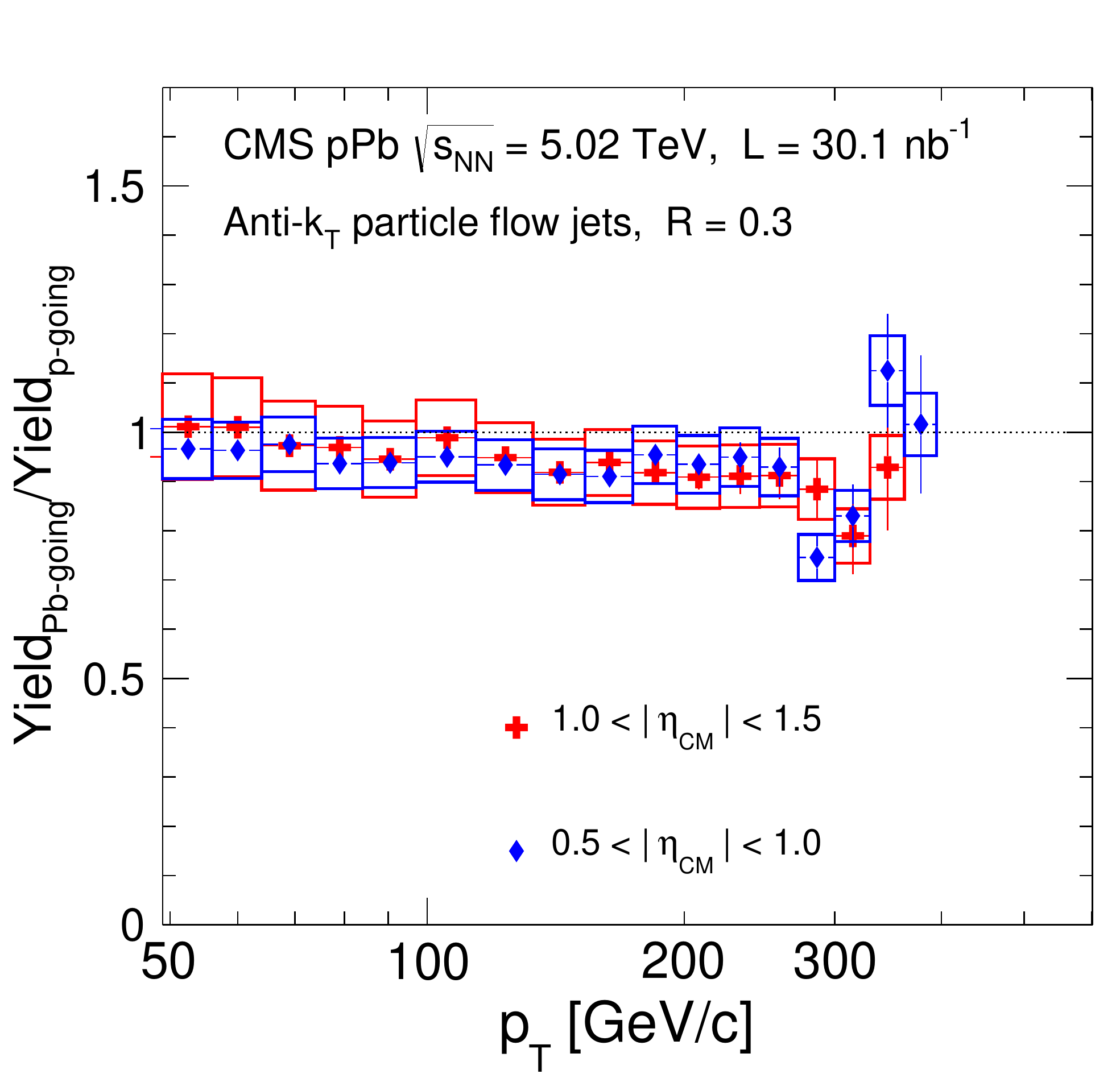}
\caption{\label{fig:Asymmetry}
Inclusive jet asymmetry as a function of jet \pt for $0.5<\abs{\eta_\mathrm{CM}}<1.0$ and $1.0<\abs{\eta_\mathrm{CM}}<1.5$. The asymmetry is calculated as the ratio between the jet yields at negative pseudorapidity (Pb beam direction) and positive pseudorapidity  (proton-going side). The vertical bars represent the statistical uncertainties and the open boxes represent the systematic ones. }
\end{figure}

The forward-backward asymmetry of the jet production
is evaluated by taking the ratio between the jet yields in the Pb-going and the proton-going directions for two pseudorapidity intervals: $0.5<\abs{\eta_\mathrm{CM}}<1.0$ and $1.0<\abs{\eta_\mathrm{CM}}<1.5$. The results are shown in Fig.~\ref{fig:Asymmetry} as a function of jet \pt. There is no significant asymmetry observed in the jet production within the covered pseudorapidity range, although a small effect at high \pt cannot be excluded with the present systematic uncertainties. The modifications in the nPDFs, if present,  are of similar magnitude in the $x$ ranges covered by the measurements in the forward and backward  directions. This result is similar to the findings from the CMS charged-hadron measurements at high \pt~\cite{Khachatryan:2015xaa}.

\begin{figure}[hpbt]
\centering
	\includegraphics[width=0.48\textwidth]{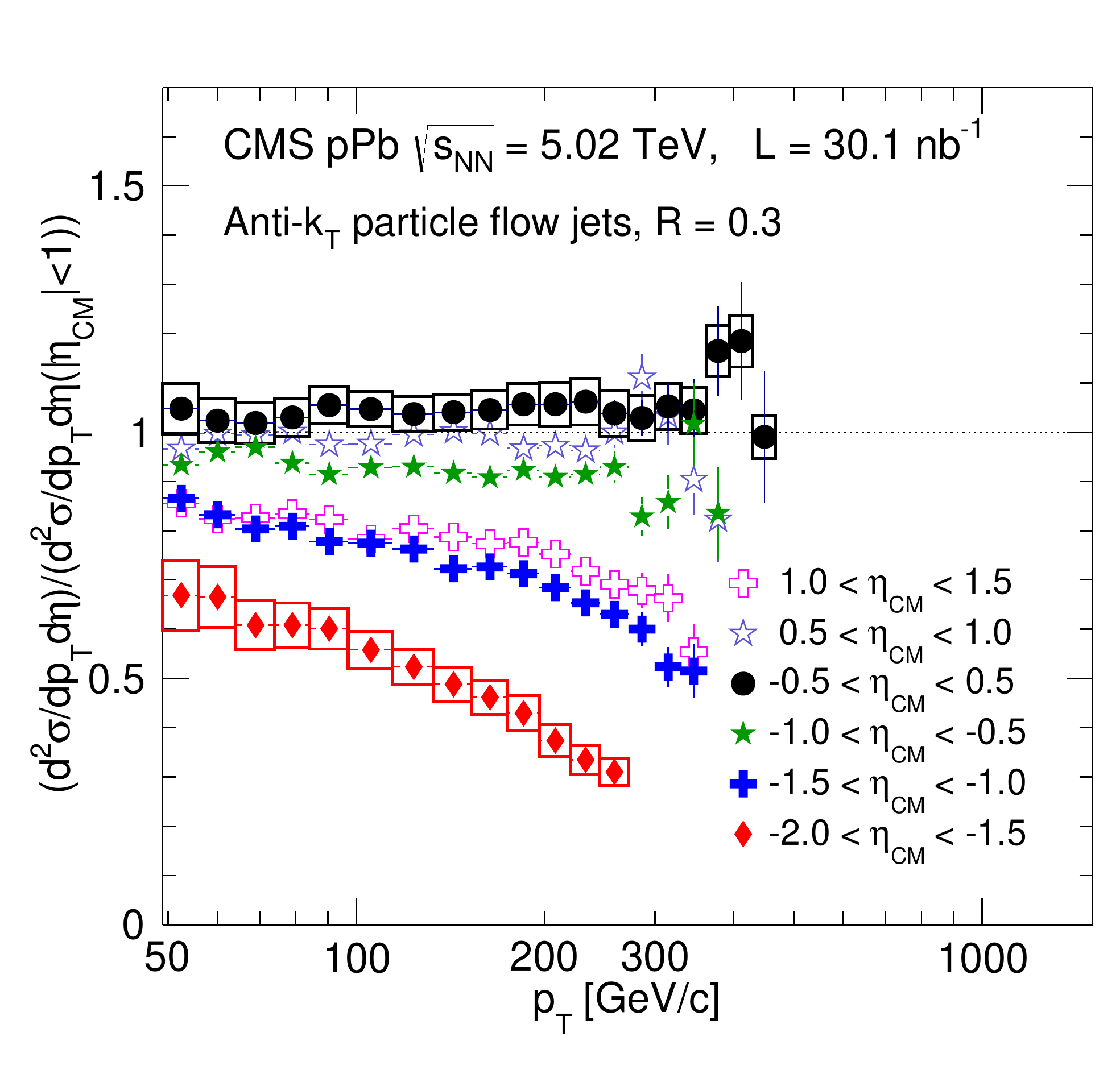}	
 	\includegraphics[width=0.48\textwidth]{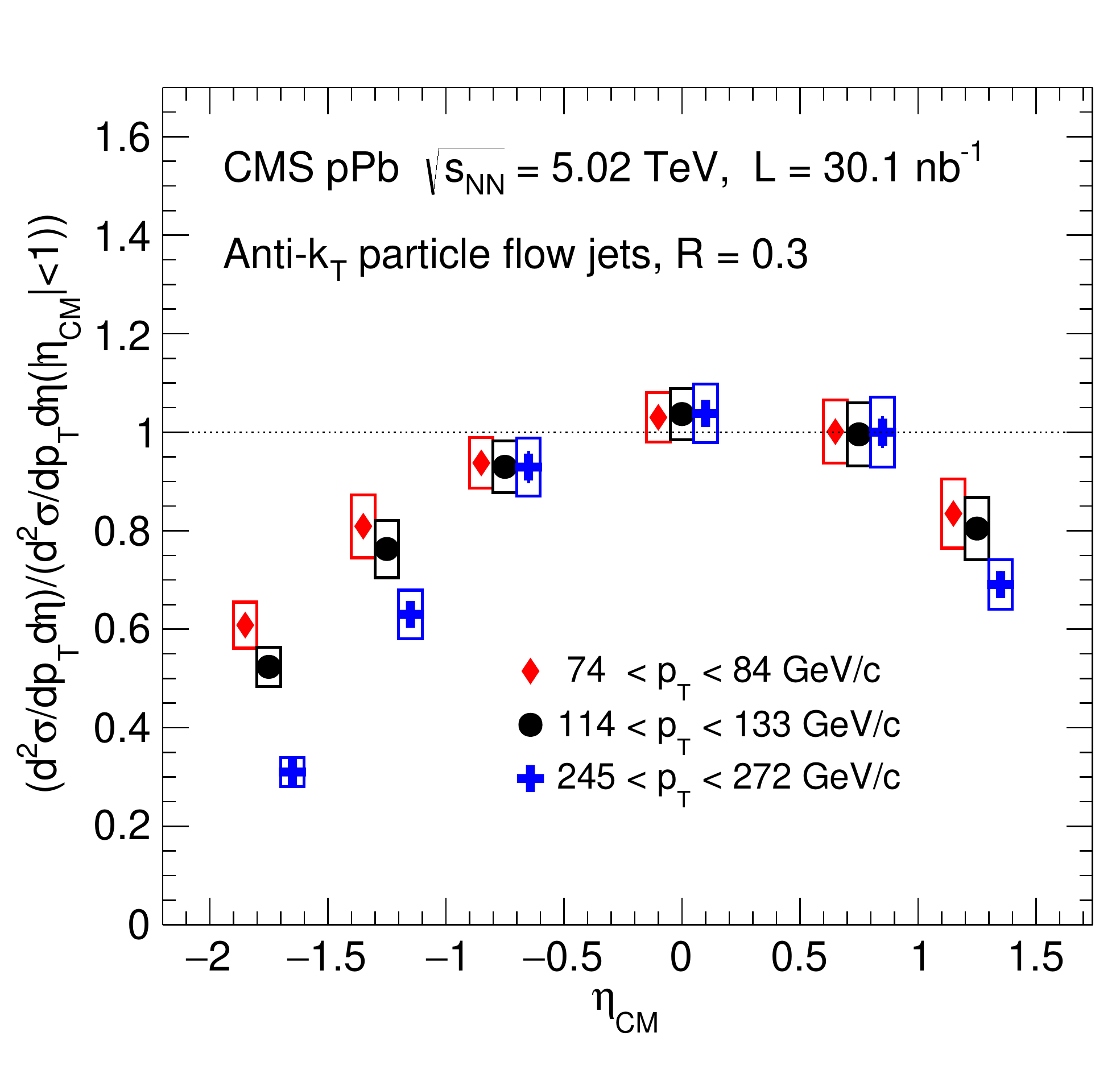}
\caption{\label{fig:ForwardMidvsEta}
\cmsLeft: Inclusive jet cross section in \pPb collisions as a function of jet \pt normalized to the production at mid-rapidity ($\abs{\eta_\mathrm{CM}}<1$) for six $\eta_\mathrm{CM}$ intervals. The vertical bars represent the statistical uncertainties. The systematic uncertainties at mid-rapidity and in the most backward pseudorapidity are shown with open boxes. The uncertainties in the other pseudorapidity ranges have similar magnitude.
\cmsRight: Inclusive jet cross section in \pPb collisions as a function of $\eta_\mathrm{CM}$  normalized to the cross section at $\abs{\eta_\mathrm{CM}}<1$, for three jet \pt ranges.
The open boxes represent the systematic uncertainties. The data points are shifted in pseudorapidity  to enhance the visibility. The $\eta_\mathrm{CM}$~bin boundaries are as specified in the \cmsleft  panel. The statistical uncertainties are smaller than the symbols.}
\end{figure}

The  evolution of the jet spectra with pseudorapidity can also be studied by normalizing each spectrum to the one obtained in the mid-rapidity range ($\abs{\eta_\mathrm{CM}}<1$).  The normalized jet cross section distributions are shown in the \cmsleft panel of Fig.~\ref{fig:ForwardMidvsEta}. In the \cmsright panel of  Fig.~\ref{fig:ForwardMidvsEta} we examine the  pseudorapidity dependence in the normalized  jet cross sections in three fixed \pt bins. The data points are offset for visibility. No significant pseudorapidity asymmetry is observed as can also be seen by comparing the open and closed stars or open and closed crosses in the \cmsleft  panel.  The jet spectra become softer away from the mid-rapidity region, and the pseudorapidity distributions become narrower with increasing jet \pt as a result of the softening of the distributions at forward and backward pseudorapidity.

\begin{figure*}[hpbt]
\centering
\includegraphics[width=1\textwidth]{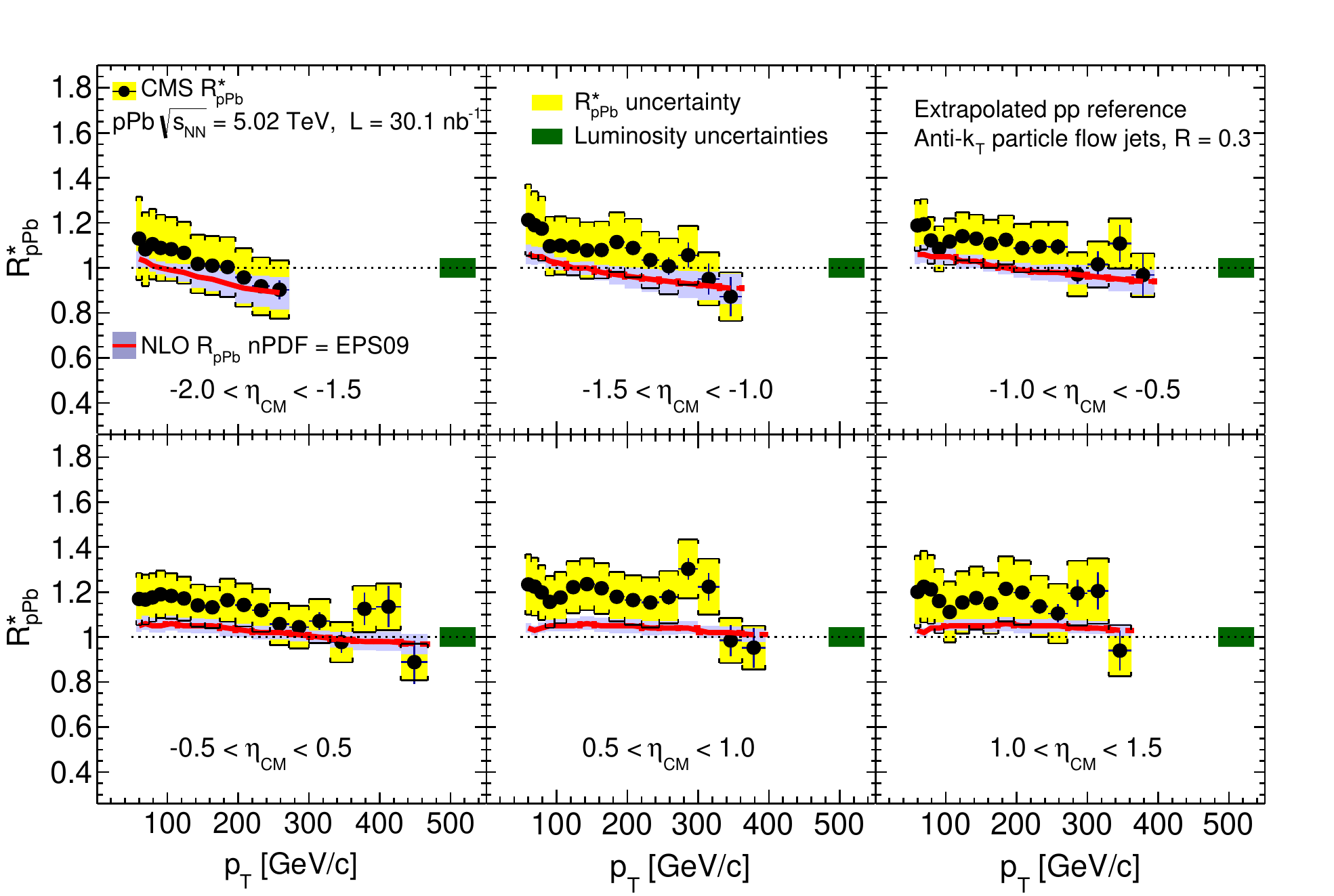}
\caption{\label{fig:JetRpPb}
Inclusive jet nuclear modification factor $R^{\ast}_\pPb$ as a function of jet \pt in
$\sqrtsNN=5.02$\TeV \pPb collisions, using a \pp reference extrapolated from previous measurements~\cite{Chatrchyan:2014gia} at $\sqrt{s} = 7 $\TeV. The vertical bars represent the statistical uncertainties, and the open boxes represent the systematic ones. The filled rectangular boxes around $R^{\ast}_\pPb = 1$ represent the  luminosity uncertainties in the \pPb and \pp measurements. The CMS measurements are compared to a NLO pQCD calculation~\cite{TheoryRpA} that is based on the EPS09 nPDFs~\cite{Eskola:2009uj}. The  theoretical calculations are shown with solid lines, and the shaded bands around them represent the  theoretical  uncertainties. }
\end{figure*}

The inclusive jet nuclear modification factors $R^{\ast}_\pPb$ as a function of jet \pt are shown in  Fig.~\ref{fig:JetRpPb} for six center-of-mass pseudorapidity bins, along with an NLO perturbative QCD (pQCD) calculation~\cite{TheoryRpA} using the EPS09 nPDFs~\cite{Eskola:2009uj}. For most of the  measured \pt and   $\eta_\mathrm{CM}$ ranges, the experimental $R^{\ast}_\pPb$  values are systematically above the theoretical prediction. However, this difference is not significant, given the size of the systematic uncertainties and the fact that they are strongly correlated in \pt.
The $R^{\ast}_\pPb$ values are approximately independent of \pt. In the theoretical prediction there is a decrease in $R_\pPb$ with \pt in the backward pseudorapidity region, which is associated with the onset of the EMC effect at high values of  $x$ in the Pb nucleus.  In the range of \pt where the measurements probe the anti-shadowing region,  the $R^{\ast}_\pPb$ values show a hint of an enhancement with respect to the \pp reference, e.g. for $\abs{\eta_\mathrm{CM}}<0.5$ and  $56< \pt < 300$\GeVc ,  $R^{\ast}_\pPb=1.17\pm0.01\stat\pm0.12\syst$.  This enhancement  is smaller than the one observed in the charged-hadron measurement~\cite{Khachatryan:2015xaa} and closer to the theoretical prediction.  Direct measurements of the jet and charged-hadron reference spectra in \pp collisions at $\sqrt{s} = 5 $\TeV are needed to reduce the systematic uncertainties in the measurements of the nuclear modification factors and provide better constraints to the theory.

\begin{figure}[hpbt]
\centering
   \includegraphics[width=0.5\textwidth]{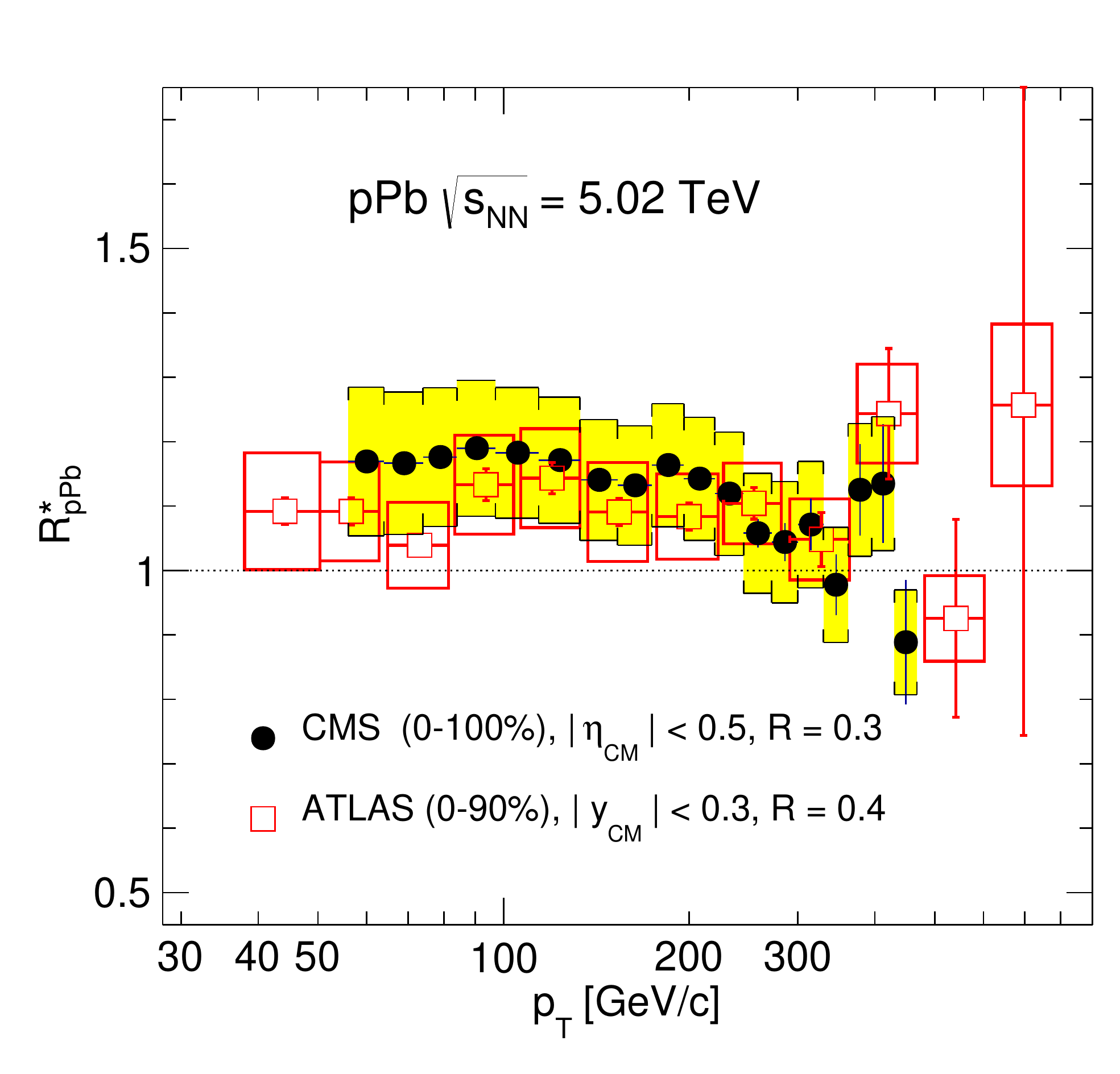}
\caption{\label{fig:ATLASComp}
Inclusive jet $R^{\ast}_\pPb$ integrated over centrality and in the  $\abs{ \eta_\mathrm{CM} }<  0.5$ range for anti-\kt jets with distance parameter $R=0.3$ from this work, compared to  ATLAS results~\cite{ATLAS:2014cpa} at  $\abs{y_\mathrm{CM} }< $ 0.3 for the 0--90\% most central collisions with distance parameter $R = 0.4$. The vertical bars show the statistical uncertainties, and the open boxes represent the systematic uncertainties.}
\end{figure}

The results of the jet $R^{\ast}_\pPb$ measurements presented here are consistent  with those reported by the ATLAS collaboration~\cite{ATLAS:2014cpa}. In Fig.~\ref{fig:ATLASComp} we compare our results to the ATLAS measurement at mid-rapidity, $\abs{y_\mathrm{CM}} < $ 0.3, for the 0--90\% most central collisions, performed using a distance parameter $R = 0.4$. Although the event selections and the jet reconstruction are not exactly the same in the two measurements, the results are in good agreement.

\section{Summary}
\label{sec:summary}

The inclusive jet spectra and nuclear modification factors in  \pPb collisions at $\sqrtsNN=5.02$\TeV have been measured. The data, corresponding to an integrated luminosity of 30.1\nbinv, were collected by the CMS experiment in 2013. The jet transverse momentum spectra were measured for $ \pt > 56$\GeVc in six pseudorapidity intervals covering  the range $ -2 <\eta_\mathrm{CM}< 1.5$ in the NN center-of-mass system. The jet spectra were found to be softer away from mid-rapidity. The jet production at forward and backward pseudorapidity were compared, and no significant asymmetry about $\eta_\mathrm{CM} = 0$  was observed in the measured kinematic range.

The differential jet cross section results were compared with extrapolated \pp reference spectra based on jet measurements in \pp collisions at $\sqrt{s} = 7$\TeV. The inclusive jet nuclear modification factors $R^{*}_\pPb$  were observed to have small enhancements compared to the reference \pp  jet spectra at low jet \pt in all $\eta_\mathrm{CM}$ ranges.  In the anti-shadowing region, for $\abs{\eta_\mathrm{CM}}<0.5$ and $56< \pt < 300$\GeVc, the value  $R^{\ast}_\pPb=1.17\pm0.01\stat\pm 0.12\syst$ was found. The $R^{*}_\pPb$ appears to be approximately independent of \pt, except in the most backward pseudorapidity range. The  $R^{\ast}_\pPb$ measurements were found to be compatible with theoretical predictions from NLO pQCD calculations that use EPS09 nPDFs.

\begin{acknowledgments}
\hyphenation{Bundes-ministerium Forschungs-gemeinschaft Forschungs-zentren} We congratulate our colleagues in the CERN accelerator departments for the excellent performance of the LHC and thank the technical and administrative staffs at CERN and at other CMS institutes for their contributions to the success of the CMS effort. In addition, we gratefully acknowledge the computing centers and personnel of the Worldwide LHC Computing Grid for delivering so effectively the computing infrastructure essential to our analyses. Finally, we acknowledge the enduring support for the construction and operation of the LHC and the CMS detector provided by the following funding agencies: the Austrian Federal Ministry of Science, Research and Economy and the Austrian Science Fund; the Belgian Fonds de la Recherche Scientifique, and Fonds voor Wetenschappelijk Onderzoek; the Brazilian Funding Agencies (CNPq, CAPES, FAPERJ, and FAPESP); the Bulgarian Ministry of Education and Science; CERN; the Chinese Academy of Sciences, Ministry of Science and Technology, and National Natural Science Foundation of China; the Colombian Funding Agency (COLCIENCIAS); the Croatian Ministry of Science, Education and Sport, and the Croatian Science Foundation; the Research Promotion Foundation, Cyprus; the Ministry of Education and Research, Estonian Research Council via IUT23-4 and IUT23-6 and European Regional Development Fund, Estonia; the Academy of Finland, Finnish Ministry of Education and Culture, and Helsinki Institute of Physics; the Institut National de Physique Nucl\'eaire et de Physique des Particules~/~CNRS, and Commissariat \`a l'\'Energie Atomique et aux \'Energies Alternatives~/~CEA, France; the Bundesministerium f\"ur Bildung und Forschung, Deutsche Forschungsgemeinschaft, and Helmholtz-Gemeinschaft Deutscher Forschungszentren, Germany; the General Secretariat for Research and Technology, Greece; the National Scientific Research Foundation, and National Innovation Office, Hungary; the Department of Atomic Energy and the Department of Science and Technology, India; the Institute for Studies in Theoretical Physics and Mathematics, Iran; the Science Foundation, Ireland; the Istituto Nazionale di Fisica Nucleare, Italy; the Ministry of Science, ICT and Future Planning, and National Research Foundation (NRF), Republic of Korea; the Lithuanian Academy of Sciences; the Ministry of Education, and University of Malaya (Malaysia); the Mexican Funding Agencies (CINVESTAV, CONACYT, SEP, and UASLP-FAI); the Ministry of Business, Innovation and Employment, New Zealand; the Pakistan Atomic Energy Commission; the Ministry of Science and Higher Education and the National Science Centre, Poland; the Funda\c{c}\~ao para a Ci\^encia e a Tecnologia, Portugal; JINR, Dubna; the Ministry of Education and Science of the Russian Federation, the Federal Agency of Atomic Energy of the Russian Federation, Russian Academy of Sciences, and the Russian Foundation for Basic Research; the Ministry of Education, Science and Technological Development of Serbia; the Secretar\'{\i}a de Estado de Investigaci\'on, Desarrollo e Innovaci\'on and Programa Consolider-Ingenio 2010, Spain; the Swiss Funding Agencies (ETH Board, ETH Zurich, PSI, SNF, UniZH, Canton Zurich, and SER); the Ministry of Science and Technology, Taipei; the Thailand Center of Excellence in Physics, the Institute for the Promotion of Teaching Science and Technology of Thailand, Special Task Force for Activating Research and the National Science and Technology Development Agency of Thailand; the Scientific and Technical Research Council of Turkey, and Turkish Atomic Energy Authority; the National Academy of Sciences of Ukraine, and State Fund for Fundamental Researches, Ukraine; the Science and Technology Facilities Council, UK; the US Department of Energy, and the US National Science Foundation.

Individuals have received support from the Marie-Curie program and the European Research Council and EPLANET (European Union); the Leventis Foundation; the A. P. Sloan Foundation; the Alexander von Humboldt Foundation; the Belgian Federal Science Policy Office; the Fonds pour la Formation \`a la Recherche dans l'Industrie et dans l'Agriculture (FRIA-Belgium); the Agentschap voor Innovatie door Wetenschap en Technologie (IWT-Belgium); the Ministry of Education, Youth and Sports (MEYS) of the Czech Republic; the Council of Science and Industrial Research, India; the HOMING PLUS program of the Foundation for Polish Science, cofinanced from European Union, Regional Development Fund; the OPUS program of the National Science Center (Poland); the Compagnia di San Paolo (Torino); MIUR project 20108T4XTM (Italy); the Thalis and Aristeia programs cofinanced by EU-ESF and the Greek NSRF; the National Priorities Research Program by Qatar National Research Fund; the Rachadapisek Sompot Fund for Postdoctoral Fellowship, Chulalongkorn University (Thailand); the Chulalongkorn Academic into Its 2nd Century Project Advancement Project (Thailand); and the Welch Foundation, contract C-1845.
\end{acknowledgments}

\bibliography{auto_generated}
\cleardoublepage \appendix\section{The CMS Collaboration \label{app:collab}}\begin{sloppypar}\hyphenpenalty=5000\widowpenalty=500\clubpenalty=5000\input{HIN-14-001-authorlist.tex}\end{sloppypar}
\end{document}

%% file: HIN-14-001-authorlist.tex
\textbf{Yerevan Physics Institute,  Yerevan,  Armenia}\\*[0pt]
V.~Khachatryan, A.M.~Sirunyan, A.~Tumasyan
\vskip\cmsinstskip
\textbf{Institut f\"{u}r Hochenergiephysik der OeAW,  Wien,  Austria}\\*[0pt]
W.~Adam, E.~Asilar, T.~Bergauer, J.~Brandstetter, E.~Brondolin, M.~Dragicevic, J.~Er\"{o}, M.~Flechl, M.~Friedl, R.~Fr\"{u}hwirth\cmsAuthorMark{1}, V.M.~Ghete, C.~Hartl, N.~H\"{o}rmann, J.~Hrubec, M.~Jeitler\cmsAuthorMark{1}, V.~Kn\"{u}nz, A.~K\"{o}nig, M.~Krammer\cmsAuthorMark{1}, I.~Kr\"{a}tschmer, D.~Liko, T.~Matsushita, I.~Mikulec, D.~Rabady\cmsAuthorMark{2}, N.~Rad, B.~Rahbaran, H.~Rohringer, J.~Schieck\cmsAuthorMark{1}, R.~Sch\"{o}fbeck, J.~Strauss, W.~Treberer-Treberspurg, W.~Waltenberger, C.-E.~Wulz\cmsAuthorMark{1}
% ...
\vskip\cmsinstskip
% ...
1:~~Also at Vienna University of Technology, Vienna, Austria\\
2:~~Also at CERN, European Organization for Nuclear Research, Geneva, Switzerland\\
% ...
